\documentclass[manyauthors]{fundam}
\usepackage{url} 
\usepackage[ruled,lined]{algorithm2e}
\usepackage{graphicx}
\usepackage{amsfonts}

\newtheorem{thm}{Theorem}

\newtheorem{defn}[thm]{Definition}
\newtheorem{rmk}[thm]{Remark}

\newcommand{\Z}{\mathbb{Z}}

\newcommand{\pk}{\mathit{pk}}
\newcommand{\sk}{\mathit{sk}}
\newcommand{\param}{\mathit{param}}
\newcommand{\keygen}{\mathrm{KeyGen}}
\newcommand{\setup}{\mathrm{Setup}}
\newcommand{\noisyenc}{\mathrm{NoisyEnc}}
\newcommand{\aggre}{\mathrm{Aggre}}
\newcommand{\aggredec}{\mathrm{AggreDec}}

\newcommand{\decrypt}{\mathrm{Decrypt}}

\newcommand{\id}{\mathit{ID}}
\newcommand{\esp}{\mathit{esp}}
\newcommand{\CT}{\mathit{CT}}
\newcommand{\ct}{\mathit{ct}}

\begin{document}

\setcounter{page}{91}
\publyear{22}
\papernumber{2143}
\volume{188}
\issue{2}

   \finalVersionForARXIV

\title{On Insecure Uses of BGN for  Privacy Preserving\\
                                         Data Aggregation Protocols}

\author{Hyang-Sook Lee\\
Department of Mathematics\\
Ewha Womans University \\ Seoul, Republic of Korea\\
hsl{@}ewha.ac.kr
\and Seongan Lim\thanks{Address for correspondence: Department of Mathematics, Inha University, Incheon,
                                     Republic of Korea. \newline \newline
                    \vspace*{-6mm}{\scriptsize{Received April 2022; \ accepted December  2022.}}},\  Ikkwon Yie\\
Department of Mathematics \\
Inha University \\
Incheon, Republic of Korea\\
seongannym{@}inha.ac.kr,\; ikyie{@}inha.ac.kr
\and Aaram Yun \\
Department of Cyber Security \\
Ewha Womans University \\
Seoul, Republic of Korea\\
aaramyun{@}ewha.ac.kr }
\maketitle

\runninghead{H-S. Lee et al.}{On Insecure Uses of BGN for Privacy Preserving Data Aggregation Protocols}

\begin{abstract}
The notion of aggregator oblivious (AO) security for privacy preserving data aggregation was formalized with a specific construction of AO-secure blinding technique over a cyclic group by Shi et al.\ Some of proposals of data aggregation protocols use the blinding technique of Shi et al.\ for BGN cryptosystem, an additive homomorphic encryption. Previously, there have been some security analysis on some of BGN based data aggregation protocols in the context of integrity or authenticity of data. Even with such security analysis, the BGN cryptosystem has been a popular building block of privacy preserving data aggregation protocol. In this paper, we study the privacy issues in the blinding technique of Shi et al.\ used for BGN cryptosystem. We show that the blinding techniques for the BGN cryptosystem used in several protocols are not privacy preserving against the recipient, the decryptor. Our analysis is based on the fact that the BGN cryptosystem uses a pairing $e:G\times G\rightarrow G_T$ and the existence of the pairing makes the DDH problem on $G$ easy to solve. We also suggest how to prevent such privacy leakage in the blinding technique of Shi et al.\ used for BGN cryptosystem.
\end{abstract}


\section{Introduction}
An additive homomorphic encryption (AHE) allows users to aggregate the data over ciphertexts, without decrypting them, and it is one of important cryptographic primitives for privacy preserving security services including data aggregation protocol. Data aggregation protocol is a useful mechanism that improves the quality of smart grid as IoT services. However, using homomorphic encryption alone does not support privacy preserving against the recipient who is the decryptor of the ciphertexts. In fact, preventing privacy leakage of individual data of the data aggregation protocol has been a challenging issue for its deployment. In 2011, Shi et al.\ considered a scenario where a group of participants periodically uploads encrypted values to a data aggregator such that the aggregator computes the sum of all participants' values in every time period and formalized the notion of aggregator oblivious (AO) security. The AO security is a formalization of the privacy requirement that the aggregator gets no information on the individual data other than the aggregated value. They also proposed a blinding technique for data aggregation protocol and proved its AO security.

When the blinding technique is employed for an additive homomorphic encryption, every fresh ciphertext is a noisy ciphertext in the eyes of the decryptor and the blinding technique of Shi et al.\ allows to manage the additional secret information for blinding off the noise from the aggregated noisy ciphertext. Some blinding techniques use a secret sharing scheme to share some information to manage the additional secret information for blinding off the noise from the aggregated noisy ciphertext.
We use the term noisy AHE for the data aggregation which uses the blinding technique with an AHE. In the noisy AHE, we see that the aggregator of the individual noisy ciphertexts can be anyone and one can set the aggregator of the individual noisy ciphertexts as a distinct entity (e.g. collector) from the decryptor (e.g. service provider).

If the underlying AHE is IND-CPA secure and the noisy ciphertext is a valid ciphertext of the underlying AHE, the aggregated ciphertexts as well as individual ciphertexts are indistinguishable from random, and thus the data privacy of the aggregated data as well as individual data is preserved against the entities except the decryptor. Therefore, the potential adversary for the data privacy of noisy AHE is the decryptor if the underlying AHE is IND-CPA secure and the noisy ciphertexts are valid ciphertexts of the underlying AHE.

The BGN cryptosystem is an IND-CPA secure additive homomorphic encryption which allows one multiplication over ciphertexts~\cite{BGN05}. There are several proposals for noisy AHE based on the BGN cryptosystem~\cite{FHL14,HKZVY17,LLWKR19,ZLC22}.
We note that the BGN cryptosystem uses a pairing $e:G\times G\to G_T$  and the decisional diffie-hellman problem on $G$ is easy to solve due to the existence of the pairing $e$. We also note that the blinding technique of Shi et al.\ on a cyclic group is proven to be AO secure under the hardness assumption of the decisional diffie-hellman problem on the cyclic group.  Therefore, the proven AO security of the blinding technique is not a direct consequence if it is used for a cyclic group $G$ where a pairing $e:G\times G\rightarrow G_T$ exists.

In this paper, we show that the noisy AHEs based on the BGN cryptosystem in~\cite{FHL14,HKZVY17,LLWKR19,ZLC22} do not satisfy the AO security against the decryptor. Our attacks are from the fact that the noisy ciphertexts in the noisy AHE of these data aggregation protocol belong to the group $G$ where DDH problem is easy to solve by using the pairing $e:G\times G\to G_T$.
Previously, there have been some security analysis on some of these data aggregation protocols in the context of integrity or authenticity of data~\cite{BL16,WXX18}, not of the privacy. Even after such a security analysis of the BGN based data aggregation protocol has been published, the BGN cryptosystem has been very popular building block of the noisy AHE to protect the privacy of individual data. Since the privacy itself is most crucial for privacy preserving data aggregation protocol, one should have a careful examination of privacy leakage against the noisy AHE using BGN cryptosystem due to the existence of pairing. Our analysis focuses on the data privacy of the protocol and does not use any advanced techniques or complicated concepts but use the basic definition of the pairing map. The existence of a pairing $e:G\times G\to G_T$ makes the DDH on the cyclic group $G$ easy but does not affect on the infeasibility of DDH on the cyclic group $G_T$. In the original BGN cryptosystem, making the ciphertext belonging to $G$ has two important purposes, one is to allow one multiplication over ciphertexts for BGN, and another is to make the size of the ciphertext shorter since $G$ is a subgroup of elliptic curve group. However, if one uses the original BGN cryptosystem in the frame of the noisy AHE, it is inevitable to leak private date as in our analysis. Fortunately, it is simple to make the noisy ciphertext to be in $G_T$ while the privacy preserving aggregation over the ciphertexts is possible according to the proven AO security from \cite{SCRCS11} since we may reasonably assume hardness of DDH, we present a BGN-based noisy AHE with ciphertexts in $G_T$.

The rest of the article is organized as
follows. In Section \ref{sec:prel}, we review some basics of
the AO security in data aggregation protocol and the blinding technique. In Section \ref{sec:analysis_BGN}, we point some issues of the AO security in noisy AHE and
present privacy analysis of the data aggregation protocol based on AHEs from~\cite{W17,FHL14,HKZVY17,LLWKR19,ZLC22}.
And the conclusion is in Section \ref{sec:conclusion}.

\section{Preliminaries}\label{sec:prel}
In the data aggregation protocol in the scenario where an aggregator aggregates private data from a set of participants periodically, Shi et al.\ presented a blinding technique that provides the aggregator oblivious property by using a cyclic group $G$ where the DDH (decisional Diffie-Hellman) assumption holds~\cite{SCRCS11}.

\begin{defn}[DDH problem]
For a cyclic group $G=\langle g \rangle$ of order $n$, the DDH problem on $G$ is defined as follows: for a given $(g^a, g^b, h)\in G^3$ with randomly chosen $0<a,b<n$, decide whether $h=g^{ab}$.
\end{defn}

\begin{defn}[Pairing]
For cyclic groups $G=\langle g \rangle$ and $G_T$, a pairing $e:G\times G \rightarrow G_T$ is a bilinear map the has the following properties.
\begin{itemize}
\item The map $e:G\times G \rightarrow G_T$ is efficiently computable.
\item For all $g_1, g_2\in G$ and $\alpha\in \mathbb{Z}$, it holds that
$$
e(g_1^\alpha,g_2)=e(g_1, g_2^\alpha)=e(g_1,g_2)^\alpha.
$$
\item The map $e$ is non-degenerate, that is, if $g_1, g_2\in G$ are generators of $G$, then $e(g_1, g_2)$ is a generator of $G_T$.
\end{itemize}
\end{defn}

\begin{rmk}\label{rmk}
We note that if a pairing $e:G\times G \rightarrow G_T$ exists, it is easy to solve the DDH problem on $G$ by checking the equality $e(g,h)=e(g^a, g^b)$.
\end{rmk}

The aggregator oblivious means that the aggregator is unable to learn any unintended information on individual data other than what it can deduce from its auxiliary knowledge.  In their blinding technique, the blinding factors are distributed to an aggregator and the users in the domain of the aggregator at the registration phase. The users blind the individual data using their own blinding factor before sending the individual data to the collector. Only when the aggregator collects the individual data of `all' participants for the time period in its domain, the aggregator can remove the blinding factors by using its blinding factor. As a result, the aggregator is able to get the sum of all participants' data in every time period without learning anything else.
Now, we recall the blinding technique of Shi et al.\ using a cyclic group $G$ in data aggregation protocol~\cite{SCRCS11}.

\begin{description}
\item{$\setup(1^\lambda)$}: Let $G$ be a cyclic group of prime order $p$ for which DDH is hard. Let $H:\Z\to G$ denote a hash function modelled as a random oracle. A trusted dealer chooses a random generator $g\in G$, and random secrets $s_0,...,s_n\in \Z_p$ such that $\sum_{i=0}^{n}s_i=0 \bmod{p}$. The public parameter $\param:=g$. The aggregator obtains $\sk_0=s_0$ and participant $i$ obtains $\sk_i=s_i$.
\item{$\noisyenc(\param,i,\sk_i, t, m_i)$}: For participant $i$ to encrypt a value $m_i\in \Z_p$ for the time stamp $t$, she computes the noisy ciphertext:
$ c_i=g^{m_i}H(t)^{\sk_i}$.
\item{$\aggredec(\param, \sk_0, t, c_1,...,c_n)$}: Compute $V=H(t)^{\sk_0}\prod_{i=1}^{n}c_i$. And compute $m=\log_g V$.
\end{description}

They also formalize the privacy notion of aggregator obliviousness by using the following Aggregator Oblivious (AO) security game~\cite{SCRCS11}.

\begin{itemize}
\item Setup: Challenger runs the $\setup$ algorithm, and returns the public parameters $\param$ to the adversary.
\item Queries: The adversary makes the following types of queries adaptively.
    \begin{itemize}
    \item Encryption Query: The adversary asks for a ciphertext of its chosen user $i$ and message $x_i$ at the time period $t$ to challenger and the challenger returns the corresponding ciphertext to the adversary.
    \item Compromise Query: The adversary specifies an integer $i\in \{0,\ldots, n\}$. If $i=0$, the challenger returns $\sk_0$, the secret of the aggregator to the adversary. If $i\ne 0$, then the challenger returns $\sk_i$, the secret of user $i$, to the adversary.
    \item Challenge Query: This query can be made only once throughout the game. The adversary specifies $U$, a set of uncompromised users and a time $t'$. For each $i\in U$, the adversary chooses two plaintexts $m_{0,i}$ and $m_{1,i}$ and sends them to the challenger. The challenger flips a random bit $b\in \{0,1\}$ and returns the noisy ciphertext of $m_{b,i}$ for the time period $t'$ for any user $i\in U$.
    \end{itemize}
\item The adversary outputs a guess for $b$.
\end{itemize}

\begin{defn}[Aggregator Oblivious (AO) Security)~\cite{SCRCS11}]
A Private Stream Aggregation is aggregator oblivious (AO) if no probabilistic polynomial time adversary has more than negligible advantage of with the above security game.
\end{defn}

The AO Security of the data aggregation protocol requires the indistinguishability of noisy ciphertexts of two messages chosen by the attackers even after compromising some of the entities in the domain. The aggregator is supposed to know the aggregated data when the aggregator aggregates all the data of the given time period in its domain, but none of uncompromised individual data. In~\cite{SCRCS11}, they proved that the above blinding technique defined over a cyclic group $G$ provides the AO security when the DDH problem on the underlying cyclic group $G$ is hard.

\section{Our Privacy analysis}\label{sec:analysis_BGN}

An additive homomorphic encryption scheme is one of most commonly used primitives in secure data aggregation protocols to provide the data privacy against outside attackers.

\subsection{Some Privacy issues of AHE based data aggregation}

If the individual data is encrypted under the public key of the recipient, it does not protect individual data privacy against the decryptor, who can decrypt the individual ciphertext and recover the individual data. However, the following data aggregation protocol uses an additive homomorphic IBE and the individual data is encrypted under the identity of the recipient.

\medskip
The identity-based data aggregation protocol presented by Wang~\cite{W17} uses an additive homomorphic IBE for privacy protection of the data aggregation protocol. For a detailed description of Wang's protocol, refer to~\cite{W17}, and we explain the contents of an individual ciphertext and how the ESP (Electricity Service Provider, decryptor) gets the individual data from the ciphertext. Wang's data aggregation protocol uses a pairing $e:G\times G\to G_T$ where $G=\langle g\rangle, \; G_T=\langle g_t \rangle$ are cyclic groups of a prime order. The encryption key of the ESP is $W=e(H(\id_\esp), g^x)$ and the private key of the ESP is $d_{\id_\esp}=H(\id_\esp)^x$. The $i$th smart grid device computes a ciphertext $\CT_i=(g^{r_i}, g_t^{m_i}\cdot W^{r_i})=(X_i, Y_i)\in G\times G_T$. Every smart grid device sends $\CT_i$ with some authentication data to the collector of the domain and the collector sends the aggregated data to the ESP. However, when ESP sees the individual ciphertext $\CT_i$, ESP can compute $m_i$ as follows:
\begin{itemize}
\item Compute $A=e(d_{\id_\esp},X_i)=W^{r_i}$, this is because $e(d_{\id_\esp},X_i )=e(H(\id_\esp)^{x}, g^{r_i})=e(H(\id_\esp)^{r_i}, g^x)$,
\item Compute $B=\frac{Y_i}{A}=g_t^{m_i}$, and then compute $m_i=\log_{g_t}B$.
\end{itemize}
As we have seen from Wang's protocol, AHE alone does not protect individual data privacy against the decryptor. Since the decryptor of the ciphertext is a potential adversary of the data privacy of the data aggregation protocol, Wang's data aggregation protocol is not privacy preserving.

To protect the individual data privacy from the decryptor of the underlying AHE, the blinding technique of Shi et al.\ can be used. Some blinding techniques use a secret sharing scheme to share some information to manage the additional secret information for blinding off the noise from the aggregated noisy ciphertext. We call noisy AHE for the data aggregation that uses an IND-CPA secure AHE where the individual ciphertexts are noisy by using the blinding technique.

\medskip
In the noisy AHE, each participant computes a noisy ciphertext periodically and data aggregation is performed over the noisy ciphertexts in the time period and then the decryptor decrypts the ciphertext of the aggregated data.
One might expect that noisy AHE is privacy preserving against outsiders directly from the IND-CPA security of the underlying AHE and it is privacy preserving against the decryptor directly from the AO security of the blinding technique of Shi et al.\  However, we point that one has to check the following for the noisy AHE to enjoy the proven IND-CPA security of the underlying AHE and AO security of the blinding technique.
\begin{itemize}
\item The IND-CPA security of the underlying AHE: one has to check that the noisy ciphertexts are valid ciphertexts under the public key of the recipient of the underlying AHE. If the noisy ciphertexts are valid ciphertext, the IND-CPA security of the underlying AHE assures the indistinguishability the noisy ciphertext from random in the eyes of all the entities except the decryptor.
\item The AO security against the decryptor: In the noisy AHE, aggregation is performed over ciphertexts. However, the decryptor can also compute the masked individual data by decrypting the noisy ciphertexts  and the aggregated plain data by using its decryption key. Therefore, one has to check the AO-insecurity of noisy AHE against the decryptor who can see not only the noisy ciphertext but also the masked individual data and the aggregated plain data.
\end{itemize}

Previously, some results of the analysis on these data aggregation protocols in the context of integrity or authenticity of data are presented~\cite{BL16,WXX18}. Our study focuses on the data privacy of the noisy AHE.

\subsection{Privacy analysis of some \textit{noisy BGN}s}

Now we show that some of the noisy AHEs based on the BGN cryptosystem are AO-insecure, even though the BGN cryptosystem is IND-CPA secure.
The BGN cryptosystem is an AHE which uses a pairing $e: G\times G\to G_T$ where $G$ and $G_T $ are cyclic group of composite order $N=pq$ under the assumption that it is infeasible to factor $N$. The IND-CPA security of the BGN cryptosystem relies on the subgroup decision problem for $G$ and $G_T$~\cite{BGN05}. The public key of the BGN cryptosystem is $\pk=(g, h)$ and the private key is $\sk=q$, where $g, h\in G$ such that $G=\langle g\rangle$ and the order of $h$ is $q$. The ciphertext $C$ of a plaintext $m$ is $C=g^mh^r\in G$ for a uniformly chosen $r\in \mathbb{Z}_q$. It is important to note that the DDH problem on $G$ is easy to solve due to the existence of the pairing $e:G\times G\to G_T$. Therefore, the proof of Shi et al.\ for the AO security of the blinding technique does not work for this cyclic group $G$. Nonetheless, several data aggregation protocol using the noisy AHEs based on the BGN cryptosystem have been proposed.

We discuss the possible privacy leakage of the noisy AHEs based on the BGN cryptosystem in~\cite{FHL14,HKZVY17,LLWKR19,ZLC22}. In~\cite{HKZVY17}, the authors describe the BGN cryptosystem incorrectly.
The noisy BGN cryptosystems in the protocols from \cite{FHL14,LLWKR19,ZLC22} use the blinding technique of Shi et al.\  We recall the Setup and NoisyEnc of the blinding techniques from~\cite{FHL14,LLWKR19,ZLC22} in the following.

\begin{description}
\itemsep=0.9pt
\item{$\keygen(1^\lambda)$}: On input security parameter $1^\lambda$, KeyGen performs as follows:\vspace*{-1mm}
    \begin{itemize}
    \item Generates a bilinear pairing $e:G\times G\to G_T$ where $G$ is a cyclic group of composite order $N=pq$.
    \item Chooses random generators $f,g,u\in G$ and computes $h=u^p$
    \item Selects a cryptographically secure hash function $H:\{0,1\}^*\to G$
    \item Outputs
        \begin{description}
        \item {In~\cite{FHL14}}: $\pk=(e, N, g, h)$ and $\sk=q$
        \item {In~\cite{LLWKR19,ZLC22}}: $\pk=(e, N, f, g, h)$ and $\sk=q$
        \end{description}
    \end{itemize}
\item{$\setup(\pk=(e:G\times G\to G_T, N, g, h))$}: A trusted dealer chooses random secrets $s_0,...,s_n\in \mathbb{Z}_N$ such that $\sum_{i=0}^{n}s_i=0 \bmod{N}$. The trusted dealer sends $\sk_0=s_0$ to the aggregator and sends $\sk_i=s_i$ for each participant $i$ securely.
\item{$\noisyenc(\param, \sk_i, t, m_i)$}: For participant $i$ to encrypt a small enough value $m_i\in \mathbb{Z}$ for which the DLP is easy with respect to the exponent $m_i$ at the time stamp $t$, NoisyEnc computes the noisy ciphertext $\CT_i$. The noisy ciphertext of~\cite{FHL14} and~\cite{LLWKR19} composed differently which are as the following:
    \begin{itemize}
    \item In~\cite{FHL14}: $\CT_i=g^{m_i}(H(t)h^{r_i})^{s_i}$ for uniformly chosen $r_i$ from $\mathbb{Z}_N$.
    \item In~\cite{LLWKR19,ZLC22}: $\CT_i=g^{m_i}h^{r_i}f^{s_i}$ for uniformly chosen $r_i$ from $\mathbb{Z}_N$
    \end{itemize}
\item{$\aggre(\param, \sk_0=s_0, \CT_1,...,\CT_n)$}: The aggregator computes the aggregated ciphertext $V$.
    \begin{itemize}
    \itemsep=0.85pt
    \item In~\cite{FHL14}: $$V=H(t)^{s_0}\prod_{i=1}^{n}\CT_i (=g^{\sum_{i=1}^{n}m_i}h^{\sum_{i=1}^{n} r_is_i})$$
    \item In~\cite{LLWKR19,ZLC22}: $$V=f^{s_0}\prod_{i=1}^{n}\CT_i(=g^{\sum_{i=1}^{n}m_i}h^{\sum_{i=1}^{n} r_i})$$
    \end{itemize}
\item{$\decrypt(\sk=q, V)$}: The owner of the private key $\sk$ computes the aggregated data by $m=\log_{g^q} V^q (=\log_{g^q} (g^q)^{\sum_{i=1}^{n}m_i})$.
\end{description}

We note that the noisy ciphertexts of each scheme are valid ciphertexts of the BGN cryptosystem under the public key of the recipient:
\begin{itemize}
\item In~\cite{FHL14}: $\CT_i=g^{m_i}(H(t)h^{r_i})^{s_i}= g^{m_i+k_t s_i} h^{r_is_i}$, which is a valid ciphertext of $m_i+k_t s_i$ for $H(t)=g^{k_t}$.
\item In~\cite{LLWKR19,ZLC22}: $\CT_i=g^{m_i}h^{r_i}f^{s_i}= g^{m_i+k' s_i} h^{r_i}$, which is a valid ciphertext of $m_i+k'_t s_i$ for $f=g^{k'}$.
\end{itemize}
Therefore, the AO security for each scheme against the adversaries other than the recipient (the decryptor) is assured from the IND-CPA security of the BGN cryptosystem.
We discuss the privacy leakage from the noisy ciphertexts against the recipient who knows the private key $\sk$ for the data aggregation schemes in~\cite{FHL14} and~\cite{LLWKR19,ZLC22} separately.

\medskip
First, we consider the noisy ciphertext $\CT_i=g^{m_i}h^{r_i}f^{s_i}\in G$  from~\cite{LLWKR19,ZLC22}.
In this scheme, the authors considered that the aggregator and the owner of the private key $\sk=q$ (recipient) are distinct and thus the aggregated ciphertext $V$ should be sent from the aggregator to the recipient through a public channel. We note that the actual blinding factor $(f^q)^{s_i}$ is fixed for all the ciphertexts computed by the user $i$. Therefore, it definitely leaks information on $m_i$ to the owner of the private key $\sk=q$. For example, suppose that two ciphertexts $\CT_i=g^{m_i}h^{r_i}f^{s_i}$ and $\CT'_i=g^{m'_i}h^{r'_i}f^{s_i}$ computed by user $i$ are given. The recipient computes $(\CT_i)^q=(g^q)^{m_i}(f^q)^{s_i}$ and $(\CT'_i)^q=(g^q)^{m'_i}(f^q)^{s_i}$, for which the randomness of the ciphertext has been removed, and thus the recipient knows whether $m_i=m_i'$ by checking whether
 $$
 (\CT_i)^q= (\CT'_i)^q
 $$

 This means that the recipient always succeeds the AO security game without comprising any other participants and the data aggregation protocol from~\cite{LLWKR19,ZLC22} violates the aggregator oblivious security.

\medskip
Now we consider the noisy ciphertext $\CT_i=g^{m_i}(H(t)h^{r_i})^{s_i}\in G$  from~\cite{FHL14}. In this scheme, the authors considered that the aggregator is the owner of the private key $\sk=q$ (recipient). The owner of  the private key $\sk=q$ can compute $\ct_i=(\CT_i)^q=(g^q)^{m_i}(H(t)^q)^{s_i}\in G$. The factor $(H(t)^q)^{s_i}$ blinds $m_i$ in $\ct_i$. We show that some information can be leaked from the existence of the pairing.
If the recipient knows the message $m_i$ of $\CT_i=g^{m_i}(H(t)h^{r_i})^{s_i}$ for a time period $t$, then the recipient knows whether the message $m_i'$ in $\CT'_i=g^{m'_i}(H(t')h^{r'_i})^{s_i}$ for the time period $t'$ is the same as $m_i$ or not by using the following steps:
\begin{itemize}
\item Compute $\ct_i=(CT_i)^q=g^{q\cdot m_i}H(t)^{q\cdot s_i}$ and $\ct'_i=(CT'_i)^q=g^{q\cdot m'_i}H(t')^{q\cdot s_i}$. Therefore, we have $\frac{\ct_i}{g^{q\cdot m_i}}=H(t)^{q\cdot s_i}$ and $\frac{\ct'_i}{g^{q\cdot m_i}}=\frac{g^{q\cdot m'_i}}{g^{q\cdot m_i}}\cdot H(t')^{q\cdot s_i}$.
\item Output $m'_i=m_i$ if $\left(H(t), H(t'), \frac{\ct_i}{g^{q\cdot m_i}}, \frac{\ct'_i}{g^{q\cdot m_i}}\right)$ is a DDH tuple, that is,
\begin{eqnarray*}
e\left(H(t),\frac{\ct'_i}{g^{q\cdot m_i}}\right)&=& e\left(H(t'),\frac{\ct_i}{g^{q\cdot m_i}}\right).
 \end{eqnarray*}
\end{itemize}
We note that $e\left(H(t),\frac{\ct'_i}{g^{q\cdot m_i}}\right)= e\left(H(t'),\frac{\ct_i}{g^{q\cdot m_i}}\right)$ is equivalent to $m'_i=m_i$ because of the following.
\begin{eqnarray*}
 e\left(H(t),\frac{\ct'_i}{g^{q\cdot m_i}}\right)&=&e\left(H(t),\quad\frac{g^{q\cdot m'_i}}{g^{q\cdot m_i}}\cdot H(t')^{q\cdot s_i}\right)\\
 &=&e\left(H(t),\frac{g^{q\cdot m'_i}}{g^{q\cdot m_i}}\right)\cdot e\left(H(t), H(t')^{q\cdot s_i}\right)\\
 e\left(H(t'),\frac{\ct_i}{g^{q\cdot m_i}}\right)&=&e(H(t'), H(t)^{q\cdot s_i}) = e\left(H(t), H(t')^{q\cdot s_i}\right)
\end{eqnarray*}
Therefore, $e\left(H(t),\frac{\ct'_i}{g^{q\cdot m_i}}\right)= e\left(H(t'),\frac{\ct_i}{g^{q\cdot m_i}}\right)$ is equivalent to $e\left(H(t),\frac{g^{q\cdot m'_i}}{g^{q\cdot m_i}}\right)=1$. From the degeneracy of the pairing $e$, it is equivalent to $\frac{g^{q\cdot m'_i}}{g^{q\cdot m_i}}= 1$, that is, $m_i=m'_i \bmod{p}$. Therefore, we have $m_i=m'_i$ since the individual data $m_i\in \mathbb{Z}$ is small enough so that the DLP with the exponent $m_i$ with the base $g^q$ is feasible in the decryption.
This means that the recipient (the decryptor) of the data aggregation protocol from~\cite{FHL14} always succeeds the AO security game without comprising any other participants.

\medskip
Both of our attacks given above use the decryptor's knowledge of the private key $\sk$  to remove the randomness parameter $r$ in the noisy ciphertext and solve a DDH problem from the pairing $e:G\times G\to G_T$. The success of our attacks are due to the fact that the transformed masked individual data $ct_i$ by using the decryption key belongs to the cyclic group $G$ where the DDH on $G$ is easy.
Since the AO security of the blinding technique of Shi et al.\ is proved if the DDH problem in the cyclic group where the noisy ciphertexts belong, one of the solutions that guarantee the AO security is that computing the ciphertexts of the BGN cryptosystem from $G_T$, not from $G$. It is known that the DDH problem is infeasible, in the current computing environment, in the cyclic group $G_T$ of the pairing $e:G\times G\to G_T$. For the BGN cryptosystem, it is relatively simple to have the ciphertexts from the cyclic group $G_T$. In such a setting, it is additive homomorphic encryption, but the feature of one homomorphic multiplication over ciphertexts of the original BGN cryptosystem is not attainable. For example, the algorithms $\keygen$, $\noisyenc$ and $\aggre$ of NoisyEnc of the blinding techniques from~\cite{FHL14,LLWKR19,ZLC22} can be replaced as follows to enjoy the proven AO security of the blinding technique of She et al.\ as well as other feature of pairing $e:G\times G\to G_T$ in the remaining blocks of the data aggregation protocol.

\begin{description}
\item{$\keygen(1^\lambda)$}: On input security parameter $1^\lambda$, KeyGen performs as follows:
    \begin{itemize}
    \item Generates a bilinear pairing $e:G\times G\to G_T$ where $G$ is a cyclic group of composite order $N=pq$.
    \item Chooses random generators $g,u\in G$ and computes $h=u^p$ and compute $\hat{g}=e(g,g), \hat{h}=e(g,h)$.
    \item Selects a cryptographically secure hash function $H:\{0,1\}^*\to G_T$
    \item Outputs $\pk=(e, N, H,  \hat{g}, \hat{h})$ and $\sk=q$
    \end{itemize}
\item{$\noisyenc(\param, \sk_i, t, m_i)$}: For participant $i$ to encrypt a small enough value $m_i\in \mathbb{Z}$ for which the DLP is easy with respect to the exponent $m_i$ at the time stamp $t$, NoisyEnc computes the noisy ciphertext $\CT_i$. Each user has $\hat{g}=e(g,g), \hat{h}=e(g,h)$ by a pre-computation.
    \begin{itemize}
    \item  $\CT_i={\hat{g}}^{m_i}(H(t){\hat{h}}^{r_i})^{s_i}\in G_T)$ for uniformly chosen $r_i$ from $\mathbb{Z}_N$.
    \end{itemize}
\item{$\aggre(\param, \sk_0=s_0, \CT_1,...,\CT_n)$}: The aggregator computes the aggregated ciphertext $V$.
   $$
    V=H(t)^{s_0}\prod_{i=1}^{n}\CT_i ({\hat{g}}^{\sum_{i=1}^{n}m_i}{\hat{h}}^{\sum_{i=1}^{n} r_is_i}))
   $$
\item{$\decrypt(\sk=q, V)$}: The owner of the private key $\sk$ computes the aggregated data by
$$m=\log_{{\hat{g}}^q} V^q (=\log_{{\hat{g}}^q} ({\hat{g}}^q)^{\sum_{i=1}^{n}m_i}).$$
\end{description}

In fact, we suggest that the additive encryption to be taken from the cyclic group $G_T$ if one uses the blinding technique of Shi et al.\ for the BGN cryptosystem which uses a pairing $e:G\times G\rightarrow G_T$.

\section{Conclusion}\label{sec:conclusion}

The BGN cryptosystem has been a popular building block of privacy preserving data aggregation protocol.
In this paper, we show that some the data aggregation protocols using the blinding technique of Shi et al.\ with the BGN cryptosystem are not privacy preserving.
Our privacy attacks use the fact that the decryptor can remove the randomness parameter $r$ in the noisy ciphertext by using its decryption key and can solve DDH problem on $G$ with the pairing. We conclude that if the additive encryption is taken from the cyclic group $G_T$ in the noisy AHE based on the BGN cryptosystem defined over a pairing $e:G\times G\rightarrow G_T$, then one can enjoy the proven AO security of the blinding technique of She et al.\ as well as other feature of pairing $e:G\times G\to G_T$.

\subsection*{Acknowledgments}
Hyang-Sook Lee was supported by the National Research Foundation of Korea (NRF) grant funded by the Korea government (MSIT)
(Grant No. 2021R1A2C1094821) and partially supported by the Basic Science Research Program through the NRF funded by the Ministry of Education (Grant No. 2019R1A6A1A11051177). Seongan Lim was supported by the National Research Foundation of Korea (NRF) grant funded by the Korea government(Grant No. 2016R1D1A1B01008562). Aaram Yun was supported by the National Research Foundation of Korea (NRF)  funded by the Ministry of Education (Grant No. 2019R1A6A1A11051177).



\begin{thebibliography}{1}
\providecommand{\url}[1]{\texttt{#1}}
\providecommand{\urlprefix}{URL }
\expandafter\ifx\csname urlstyle\endcsname\relax
  \providecommand{\doi}[1]{doi:\discretionary{}{}{}#1}\else
  \providecommand{\doi}{doi:\discretionary{}{}{}\begingroup
  \urlstyle{rm}\Url}\fi
\providecommand{\eprint}[2][]{\url{#2}}

\bibitem{BGN05}
 Boneh D,  Goh E, and Nissim K.
\newblock Evaluating 2-DNF formulas on ciphertexts.
\newblock TCC 2005.
\newblock LNCS 3378, 2005. pp.~325--342.
doi:10.1007/978-3-540-30576-7\_18.

\bibitem{BL16}
 Bao H, and Lu R.
 Comment on ``Privacy-Enhanced Data Aggregation Scheme Against Internal Attackers in Smart Grid".
 IEEE Transactions on industrial informatics, 2016. 12(1):1–5. doi:10.1109/TII.2015.2500882.

\bibitem{FHL14}
Fan C-I,  Huang S-Y, and Lai Y-L.
\newblock Privacy-Enhanced Data Aggregation Scheme Against Internal Attackers in Smart Grid.
\newblock IEEE Transactions on industrial informatics,
\newblock  2003. 10(1):666--675.

\bibitem{HKZVY17}
 He D, Kumar N,  Zeadally S,  Vinel A, and  Yang L.
\newblock Efficient and Privacy Preserving Data Aggregation Scheme for Smart Grid Against Internal Adversaries.
\newblock IEEE Transactions on smart grid,
\newblock 2017.  8(5):2411--2417.

\bibitem{LLWKR19}
 Li X,  Liu S,  Wu F,  Kumari S, and  Rodrigues J.
\newblock Privacy Preserving Data Aggregation Scheme for Mobile Edge Computing Assisted IoT Applications.
\newblock IEEE Internet of Things Journal,
\newblock 2019.  6(3):4755--4763.

\bibitem{SCRCS11}
 Shi E, Chan T-H, Chan H, Rieffel E, Chow R, and  Song D.
\newblock Privacy-Preserving Aggregation of Time-Series Data.
\newblock NDSS 2011. pp.~483--501.

\bibitem{W17}
 Wang Z.
\newblock An Identity-based data aggregation protocol for the smart grid.
\newblock IEEE Transactions on industrial informatics,
\newblock 2017.  13(5):2428--2435. doi:10.1109/TII.2017.2705218.

\bibitem{WXX18}
 Wang Z,  Xie H, and  Xu Y.
\newblock Security Analysis of an Identity-Based Data Aggregation Protocol for the Smart Grid.
\newblock ISDDC 2018.
\newblock LNCS 11317, 2018. pp.~63--73.  doi:10.1007/978-3-030-03712-3\_6.

\bibitem{ZLC22}
 Zeng Z,  Liu Y, and Chang L.
\newblock A Robust and Optional Privacy Data Aggregation Scheme for Fog-Enhanced IoT Network.
\newblock IEEE Systems Journal, 2022. pp.~1--11. doi:10.1109/JSYST.2022.3177418.
\end{thebibliography}
\end{document}